\documentstyle[aps,prl,epsfig]{revtex}

\begin{document} 
\wideabs{
\title{Dark solitons in a two-component Bose--Einstein condensate} 
\author{P. \"Ohberg and L. Santos} 
\address{Institut f\"ur Theoretische Physik, Universit\"at Hannover, 
 D-30167 Hannover,Germany} 
 
\maketitle 
 
\begin{abstract} 
The creation and interaction of dark solitons in a two-component Bose-Einstein condensate  
is investigated.  
For a miscible case, the interaction of dark solitons in different components is studied. 
Various possible scenarios are presented, 
including the formation of a soliton--soliton bound pair.
We also analyze the soliton propagation in the presence of domains, and  
show that a dark soliton can be transferred from one component to  
the other at the domain wall when it exceeds a critical velocity. 
For lower velocities 
multiple reflections within the  
domain are observed, where the soliton is evaporated and accelerated after  
each reflection until it finally escapes from the domain. 
\end{abstract} 
} 
 
The realization of Bose--Einstein Condensation (BEC) in  
weakly interacting atomic gases \cite{BEC} has opened the possibility to  
investigate nonlinear properties of atomic matter waves.  
In this respect, several remarkable results have been reported 
such as the experimental observation of four--wave mixing in BEC \cite{4WM},  
and the realization of vortices \cite{JILA,ENS}, and dark solitons in BEC  
\cite{NIST,Hannover}.
  
A dark soliton in BEC is a macroscopic excitation of the condensate 
with a corresponding positive scattering length, 
which is characterized by a local density minimum and a sharp phase gradient of the 
wavefunction at  
the position of the minimum \cite{KivLuth}. The shape of the dip does not change due to the 
balance between  
kinetic energy and repulsive atom--atom collisions.  
The recent experimental creation of dark solitons in BEC by means  
of the phase imprinting technique \cite{NIST,Hannover}, has posed several  
fundamental questions concerning the dynamics, stability and dissipation in such systems  
\cite{Busch,Andrei,Gora}. Also, the interaction of two solitons in a BEC has been experimentally  
addressed \cite{Burger2}.

In the recent years, the development of trapping techniques has allowed the creation of  
multi-component condensates. These are formed by trapping atoms in different internal  
(electronic) states \cite{JILA2,MIT}. The multicomponent BEC,  
far from being a trivial extension of the single--component one, presents novel and fundamentally  
different scenarios for its ground--state 
wavefunction \cite{Lo} and excitations \cite{Patrik}.  
In particular, it has been experimentally observed that the BEC can reach  
an equilibrium state characterized by the phase separation of the species in different domains \cite{MIT}. 
 
In the present Letter, we analyze the creation, propagation and interaction of dark solitons in a  
two--component condensate using analytical and numerical methods. 
In this more complex scenario, novel phenomena can be expected, as 
it has been already reported in the context of nonlinear optics \cite{KivLuth}. 
We show that the dynamics of the soliton interaction is completely different from  
the single--component case. In particular, two dark solitons in a single--component BEC 
always repel each other \cite{Zakharov}, whereas the opposite is true for two solitons interacting  
in the two--component case \cite{foot0}. We consider two situations. For the first one, 
the two components are miscible, and one soliton is created in each component.  
We show, both analytically and numerically, that this system presents  
nontrivial dynamics, which includes the formation of a soliton--soliton bound state.  
The properties of such a binary bound state, in particular the period of its oscillations,  
should be sensible to dissipation effects and therefore constitutes an excellent 
tool to analyze these effects. In addition, the dissipation can be studied  
in a much clearer and controllable way than in current experiments with single-component BEC,  
since the soliton--soliton interaction keeps the solitons within a 
central region of the trap of several healing lengths. 
In the second scenario, the propagation of a soliton in a two--component BEC  
with domains is considered. In particular, we demonstrate that the soliton can be  
transferred from one component to the other at the domain wall.  
Below a critical velocity the soliton is reflected, performing  
multiple oscillations inside of the corresponding domain. At each reflection the soliton is  
partially evaporated by emission of phonons, and accelerates until it eventually escapes though the wall. 
 
In the following we consider a trapped BEC with two components, where the 
dynamics takes place only in one dimension
due to the strong trap confinement in the transverse direction.
This approximation is valid if the mean--field  
interaction is smaller than the  
typical energy separation $E_{\perp}$ in the other directions \cite{foot3},  and has been successfully  
employed in the analysis  
of dark solitons in single-component condensates \cite{Hannover}. 
For sufficiently low temperatures the dynamics is well  
described by two coupled Gross--Pitaevskii equations 
\begin{eqnarray} 
i\hbar\frac{\partial}{\partial t}\psi_j(x,t)&=& \left  
\{ -\frac{\hbar^2}{2m}\frac{\partial^2}{\partial x^2}+V(x)+  
g_{jj}|\psi_j(x,t)|^2+ \right. \nonumber \\ 
&&\left. g_{jl}|\psi_l(x,t)|^2 -\mu_j 
\right \} 
\psi_j(x,t), 
\label{GPE} 
\end{eqnarray} 
where $V(x)$ is the trap potential, $\mu_j$ is the chemical potential  
of component $j$, $g_{jl}=4\pi\hbar^2a_{jl}/mS$ is  
the coupling constant between the components $j$ and $l$ with the transversal area $S$,  
$m$ is the atomic mass and $a_{jl}$ the scattering length between $j$ and $l$ ($j,l=1,2$).
 
We consider the situation in which a  
dark soliton is created in one or in both components. 
To generate dark solitons we use the well established method of phase  
imprinting \cite{phase}, although other methods \cite{Dum} could in principle be employed. 
This method consists of applying a homogeneous potential $U$ generated by the  
dipole potential  
of a far detuned laser beam to one part of the condensate wavefunction of one  
component. The  
potential is pulsed on for a time $t_p$, such that the wavefunction locally  
acquires an  
additional phase factor $\exp (iUt_p/\hbar)$. The pulse duration is chosen to  
be short compared to the minimal correlation time of the system $\hbar/\mu$,  
with $\mu=\max (\mu_1,\mu_2)$. This ensures that the effect of the light pulse is  
mainly a change in the  
phase of the condensate, whereas changes of the density during this time can  
be neglected. 
Note, however, that due to the imprinted phase at larger times an adjustment of the  
density in the condensate appears, leading to the  
formation of the dark soliton and  
also additional structures. In principle the described method can be employed  
to selectively create  
a soliton in only one of the components, or to create two different  
solitons in each one. The creation of a  
soliton in one component modifies the density in the other  
one, due to the coupling in Eq.\ (\ref{GPE}). 
But as we show below, under appropriate conditions, solitons can  
also be created in a two--component condensate. 
 
We consider first the case in which one soliton is created in each component.  
This situation can be analytically studied by employing a variational  
approach \cite{perez,kivshar}. We assume for simplicity 
the solitons are moving in a homogeneous condensate of densities $n_1=n_2=n_0$.  
This situation corresponds to a very elongated trap with equal concentrations 
of both components.   
Additionally we assume that the coupling constants are  
$g_{11}=g_{12}=g_{22}=g$. 
The latter assumption matches well the experimental conditions\cite{JILA,MIT}. 
As variational wavefunctions the single--component soliton solutions are considered
\begin{mathletters} 
\begin{eqnarray} 
&&\frac{\psi_1(x,t)}{\sqrt{n_0}}=  \frac{i\dot q}{c_s}  
-\sqrt{1-\frac{\dot q^2}{c_s^2}} 
\tanh \sqrt{1-\frac{\dot q^2}{c_s^2}}\frac{(x-q)}{l_0} ,  
\label{variat1} 
\\ 
&&\frac{\psi_2(x,t)}{\sqrt{n_0}}= \frac{-i\dot q}{c_s}  
+\sqrt{1-\frac{\dot q^2}{c_s^2}} 
\tanh \sqrt{1-\frac{\dot q^2}{c_s^2}}\frac{(x+q)}{l_0},  
\label{variat2} 
\end{eqnarray} 
\end{mathletters} 
where $\mu_1=\mu_2=\mu$, $\dot q=dq/dt$, $2 q(t)$ denotes the relative  
distance between the solitons, $c_s=\sqrt{gn_0/m}$ the  
sound velocity and $l_0=\hbar/\sqrt{gn_0m}$ is   
the coherence length for a single component.   
Eqs.\ (\ref{variat1}) and (\ref{variat2}) represents a kink-antikink situation, i.e.  
when the phase fronts of the solitons are facing each other. The  
case kink-kink, when both phase fronts are in the same direction, is also discussed below. 
The previous expressions describe a symmetric situation around $x=0$, or  
equivalently they describe  
the system in the center of mass frame. 
 
The problem of solving Eq.\  
(\ref{GPE}) can be restated as a variational  
problem \cite{perez,kivshar}, corresponding to the stationary point of the action related to the  
Lagrangian density 
\begin{eqnarray} 
{\cal L}&=&\sum_{j=1,2} \left \{ 
\frac{i\hbar}{2}\left ( 
\psi_j\frac{\partial\psi_j^{\ast}}{\partial t}- 
 \psi_j^{\ast}\frac{\partial\psi_j}{\partial t} \right ) 
-\frac{\hbar^2}{2m}\left | \frac{\partial\psi_j}{\partial x} \right |^2  
\right. \nonumber \\ 
&&\left. +\frac{g}{2}|\psi_j|^4 -\mu|\psi_j|^2 \right \} 
+g|\psi_1|^2|\psi_2|^2. 
\label{Ldens} 
\end{eqnarray} 
Our goal is to find the equation which governs the evolution of $q(t)$. In  
order to do that, we insert  
the variational Ansatz (\ref{variat1}),(\ref{variat2}) into Eq. (\ref{Ldens}), and calculate  
an effective Lagrangian $L=\int dx {\cal L}$ which becomes 
\begin{eqnarray} 
L&=&-2gn_0^2l_0\left (1-\frac{\dot q^2}{c_s^2} \right )^{3/2} 
\frac{\cosh (2b)}{\sinh^3(2b)}(\sinh (4b)- 4b) \nonumber \\ 
&&-4gn_0^2l_0\frac{\dot q^2}{c_s^2}\sqrt{1-\frac{\dot q^2}{c_s^2}}, 
\label{Lagr} 
\end{eqnarray} 
where $b(q,\dot q)=\sqrt{1-\dot q^2/c_s^2} (q/l_0)$.  
The equation for $q(t)$ is provided by the  
Euler--Lagrange equation 
$ 
\frac{d}{dt} \left ( \frac{\partial L}{\partial\dot q} \right )- 
\frac{\partial L}{\partial q}=0, 
$ 
which we have solved for different initial conditions, obtaining the trajectories in the  
phase space $(q,\dot q)$ (Fig.\ \ref{fig:1}). For small deviations from  
$(0,0)$, the system behaves periodically 
, i.e. the solitons form a bound pair (soliton molecule).  
On the other hand the free trajectories are characterized by the acceleration  
of the approaching solitons, and the deceleration of the outgoing ones. 
For $q\gtrsim 2l_0$, $L$ becomes the sum of two single-soliton  
Lagrangians and $\ddot q$ vanishes. 
For practical purposes the trajectories can be  
considered periodic  
if they cross $\dot q=0$ at $q\lesssim 2l_0$, and free otherwise. 
The free trajectories close to the periodic ones, are squeezed together at $q=0$ and 
$\dot q=\dot q_c=0.73c_s$, which constitutes the critical escape velocity.  
 
The kink-kink situation is however different. In this case the solitons are propagating in the same  
direction with velocities $\dot q_1$ and $\dot q_2$, such that in our formalism 
$2\dot q=\dot q_2-\dot q_1$. 
For this case $L$, which has already been derived in the analysis of the stability of optical vector 
dark solitons \cite{kivshar2}, takes the form of the first line of Eq. (\ref{Lagr}).    
For small velocities also bound soliton solutions appear, oscillating for small deviations 
from $(q=0,\dot q=0)$ with a frequency $\sqrt{8/15}\mu/\hbar$. However,  
contrary to the kink-antikink case, the solitons can never break the molecule. This is reflected  
in our formalism by the appearance of a singularity at the escape velocity $\dot q=c_s/\sqrt{2}$, 
which is the  
half of the sound velocity for the homogeneous  
two--component gas $c_{s,2}=\sqrt{2}c_s$. The singularity reflects the fact that the solitons 
cannot move  
in the laboratory frame faster than $c_{s,2}$. 
The special kink-kink case $(q=0,\dot q=0)$ coincides with the optical vector dark soliton solution 
\cite{foot0,kivshar2}.
 
We have studied the creation of a soliton in each component  
using phase imprinting where different initial conditions corresponding to different  
phase imprintings are obtained. We consider the case of a BEC  
with equal number of atoms in both components, $N=10^5$, in a 
box trap and with $g_{11}=g_{22}=1.05g_{12}$. This choice allows a 
homogeneous region of equal  densities for both components,  
and therefore provides a better quantitative comparison with the  
analytical model. However,  
for more general non--homogeneous situations a similar qualitative picture  
has been observed in simulations. 

We restrict ourselves here to the case kink-antikink. 
The creation of a soliton in one component perturbs the density of the other  
one. We observe
that if the phase imprinting is applied in order to create two solitons which are  
initially separated by  
distances larger than approximately $4\l_0$, the fluctuations in the densities  
prevent the formation of the solitons. Therefore, the solitons have to be  
created initially with $q(0)<2l_0$.  
The soliton in each component induces a local density increasement  
centered at the position of the soliton in the other component. In other words,  
the solitons are filled by the other component. Therefore, they become wider and  
slower than a corresponding single-component soliton.
This fact introduces some quantitative corrections to the analytical  
estimates, although such corrections are in fact small. 

Fig.\ \ref{fig:2} shows the case of an initial $\dot q=0.81c_s$.  
For such velocity, the solitons  
move periodically around $q=0$, i.e. they are forming a soliton molecule,  
as described above. 
The separation of the solitons, which can reach $4\mu$m, depends on the initial velocity
given by the phase imprinting.
We have numerically found a critical velocity $\dot q_c=0.83c_s$ at which the 
solitons become free. This velocity is in good agreement with the variational approach.  
 
Fig.\ \ref{fig:3} shows the evolution of the solitons corresponding to 
an initial velocity $\dot q=0.89c_s$ (dashed line in Fig.\ \ref{fig:3}).  
The solitons indeed move apart much slower with $\dot q=0.095c_s$.  
The closer $\dot q$ is to $\dot q_c$ the larger the soliton deceleration is  
in agreement with our analytical results.  
Therefore, the deceleration is indeed an effect of the  
soliton--soliton interaction and not a consequence of the filling of the 
 soliton by the other component.  
We have also depicted In Fig.\ \ref{fig:3} the trajectory after the reflection from the 
 box boundaries, in order to illustrate the behavior when both solitons collide.  
As predicted from our variational approach, it can be observed that the  
solitons are accelerated when approaching each other, and decelerated after crossing. 
The maximal velocity at $q=0$ is comparable to the critical velocity. 

In the last part of this Letter we analyze numerically and analytically 
the soliton propagation in separate domains. 
If the relation between the coupling constants and densities is  
appropriately chosen, separate domains of each component  
can be created \cite{MIT}.  
We consider the case in which a soliton is created in one of the components  
and move towards the domain wall. In order to illustrate the different  
possible scenarios, 
we study the situation in which $g_{12}/g_{11}=1.7$,  
$g_{22}/g_{11}=0.96$  
for different initially imprinted velocities. Both components
have equal number of atoms $N=10^5$. 
A sufficiently fast soliton will be transferred  
through the domain wall into the other component.  
However, if the velocity is sufficiently low, the soliton is reflected at the  
domain wall,  
as shown in Fig.\ \ref{fig:4}. This figure shows the case of a box  
trap with the initial soliton velocity $\dot q=0.15 c_s$.
At each reflection the soliton is partially evaporated in the form of phonons in  
the second component.  
The latter induces an acceleration until the soliton eventually escapes the  
domain. The critical escape velocity can be estimated from simple energetic  
considerations,  assuming that the soliton  
must overcome a potential barrier induced by the second component at the domain wall. 
This gives a critical velocity  
$\dot q_t=\sqrt{(g_{12}-g_{11})/4Sl_0}$, where $S$ is the  
transversal area. In the considered example, the analytical value 
$\dot q_t=0.19c_s$ is in excellent agreement with the numerical one $\dot q_t=0.16c_s$. 
When the soliton is transferred a back action of the soliton on the domain is observed. This introduces 
density fluctuations and perturbations in the domain walls, which slightly modify the 
critical velocity. The latter can produce a retrapping of the soliton in the original domain, as 
observed in Fig.\ \ref{fig:4}.

In this Letter we have shown the rich behavior of solitons in  
two--component BEC. The two components provide solutions  
such as bound solitons and the possibility to create  
extremely slowly moving ones. We have analytically studied the dynamics  
of the system with a variational approach, and determined the  
possible scenarios.  
We have finally analyzed a two--component BEC which  
contains domains, and  
showed that depending on the physical parameters a dark soliton can be either  
transferred or reflected at the domain wall.
 
Several interesting problems remain, however, open. 
Among them, we stress especially two. 
In the present Letter we have analyzed a 1D system.  
If the 1D conditions are not  
strictly fulfilled, dynamical instability is expected \cite{Andrei,Gora}.  
In the new scenario with two--component condensates the properties of  
such instability should be altered.  
A second interesting problem is given by the dissipation of the  
oscillatory motion. The two solitons radiate phonons when oscillating.  
Contrary to the case of other binary systems, the radiation will  
increase the elongation of the oscillations,  
until eventually breaking the soliton molecule, and therefore  
these systems could be an excellent probe for the  
dissipation effects \cite{foot1}. 
 
We should finally stress that the effects considered here appear for  
realistic situations and can be experimentally analyzed with the state of  
the art technology. The creation of dark solitons constitutes a well  
established technique for the case of a single-component  
BEC. We have numerically simulated the phase imprinting mechanism in a  
two--component BEC and demonstrated that this technique can also be applied  
in that situation \cite{foot2}. 
Since the solitons are indeed wider due to the presence of the second  
component,  
some of the predicted effects, as for example the appearance of a critical  
escape velocity,  
could be experimentally observed in a non--destructive way. Others, however,  
as for example  
the soliton oscillations, could require the opening of the trap, and  
subsequent condensate expansion.  
The dynamics of such expansion will be the subject of a separate investigation. 
 
We acknowledge  support from Deutsche Forschungsgemeinschaft (SFB 407),  
TMR ERBXTCT-96-002, and  
ESF PESC BEC2000+. Discussions with J. Anglin, J. Arlt, D. Hellweg, M. Kottke, M. Lewenstein, A. Sanpera,  
H. Schmaljohann and K. Sengstock are acknowledged.

\begin{figure}[ht] 
\begin{center}\ 
\epsfxsize=4.5cm 
\hspace{0mm} 
\end{center} 
\caption{Phase map of the kink--antikink relative motion.} 
\label{fig:1}  
\end{figure} 
\begin{figure}[ht] 
\begin{center}\ 
\epsfxsize=5.5cm 
\hspace{0mm} 
\end{center} 
\caption{Density of component $1$ for the kink--antikink case
with $q(0)=0$ and $\dot q(0)=0.81c_s$. Darker regions are those with less density.
Component $2$ is the mirror image of component $1$ around $x=0$.} 
\label{fig:2}  
\end{figure} 
\begin{figure}[ht] 
\begin{center}\ 
\epsfxsize=6.0cm 
\hspace{0mm} 
\end{center} 
\caption{(Left) Density of component $1$ for the kink--antikink case  
with $q(0)=0$ and $\dot q(0)=0.89c_s$. Darker regions are those with less density.
Component $2$ is the mirror image of component $1$ around $x=0$. 
The dashed line is the soliton trajectory in a single-component condensate. 
(Right) Detail of the collision region.} 
\label{fig:3}  
\end{figure} 
\begin{figure}[ht] 
\begin{center}\ 
\epsfxsize=4.0cm 
\hspace{0mm} 
\end{center} 
\caption{Interaction with a domain wall of a soliton initially created in component $1$ with 
$\dot q=0.15 c_s$.} 
\label{fig:4}  
\end{figure}


\begin{references}
\bibitem{foot0} This situation should not be confused with the interaction of two vector 
dark solitons \cite{KivLuth} which are formed by a kink in each component, 
and are centered at the same position moving as a whole. 
See e.g. \cite{kivshar2}.
\bibitem{foot1} For the case of Fig.\ \ref{fig:2}, we have observed that the molecule  
breaks apart at times of the order of $10$s, due to dynamical instability. Scattering of thermal excitations 
will presumably speed up this decay
\bibitem{foot2} If the two solitons could not be  
independently created, due to experimental limitations, there are alternative  
ways of creating soliton molecules. E.g. the same phase imprinting 
in both components, together with a displacement of one of the  
components, also leads to the same effect.
\end{references}
\end{document}